\documentclass[12pt]{article}

\usepackage{amssymb}
\usepackage{amsmath,graphicx,amsfonts,epsfig,eucal}
\usepackage{amsfonts}

\usepackage{amsthm}

\usepackage[utf8]{inputenc}
\usepackage[english]{babel}

\newtheorem{defn}{Definition}

\begin{document}
\title{On the Solutions of the Lucas-Uzawa Model}
\author{Constantin Chilarescu}
\date{}

\maketitle

\centerline{\it Laboratoire CLERSE Universit\'{e} de Lille, France}

\centerline{\it E-mail: Constantin.Chilarescu@univ-lille.fr}

\maketitle

\begin{abstract}
In a recent paper, Naz and Chaudry provided two solutions for the model of Lucas-Uzawa, via the Partial Hamiltonian Approach. The first one of these solutions coincides exactly with that determined by Chilarescu. For the second one, they claim that this is a new solution, fundamentally different than that obtained by Chilarescu. We will prove in this paper, using the existence and uniqueness theorem of nonlinear differential equations, that this is not at all true.
\end{abstract}

{\small {\bf Keywords}: Partial Hamiltonian approach, Lucas-Uzawa model, Uniqueness of solutions.}

{\small {\bf AMS Subject Classification}: 35L65; 76M60; 83C15}

\date{}
\maketitle
\section{Introduction}\label{s:1}
The model of Lucas-Uzawa is characterized by the well-known optimization problem.
\begin {defn}
The set of paths $\left\{k, h, c, u\right\}$ is called an optimal
solution if it solves the following optimization problem:
\begin {equation}\label{OP}
V_0 = \max\limits_{u,c}\int\limits_0^{\infty}\frac{c(t)^{1-\sigma}-1}{1-\sigma}e^{-\rho t}dt,
\end {equation}
\noindent subject to
\begin {equation}\label{RC}
\left\{
 \begin{array}{lll}
\dot{k}(t) = \gamma k(t)^\beta\left[u(t)h(t)\right]^{1-\beta} - \pi k(t) - c(t),\\\\
\dot{h}(t) = \delta[1 - u(t)]h(t),\\\\
k_0 = k(0),\;h_0 = h(0),
  \end{array}
  \right.
\end {equation}
\noindent where $k_0 > 0$ and $h_0 > 0$ are given, $\beta$ is the
elasticity of output with respect to physical capital, $\rho$ is a
positive discount factor, the efficiency parameters $\gamma > 0$ and
$\delta > 0$ represent the constant technological levels in the good
sector and, respectively in the education sector, $k$ is physical capital,
$h$ is human capital, $c$ is the real per-capita consumption and $u$
is the fraction of labor allocated to the production of physical capital.
$\sigma^{-1}$ represents the constant elasticity of intertemporal
substitution, and throughout this paper we suppose that $\sigma \neq \beta$.
\end {defn}
The dynamical system that drives the economy over time is given by
\begin{equation}\label{DS}
\left\{
  \begin{array}{llllll}
\frac{\dot{k}}{k} = \gamma \left(\frac{hu}{k}\right)^{1-\beta} - \pi
- \frac{c}{k}\\\\
\frac{\dot{h}}{h} = \delta(1 - u)\\\\
\frac{\dot{\lambda}}{\lambda} = \rho+\pi-\beta\gamma
\left(\frac{hu}{k}\right)^{1-\beta}\\\\
\frac{\dot{\mu}}{\mu} =
\rho-\delta\\\\
\frac{\dot{c}}{c} = -\frac{\rho+\pi}{\sigma}+\frac{\gamma\beta}{\sigma}
\left(\frac{hu}{k}\right)^{1-\beta}
\\\\
\frac{\dot{u}}{u} = \varphi-
\frac{c}{k}+\delta u,\;\varphi = \frac{(\delta+\pi)(1-\beta)}{\beta}.
\end{array}
  \right.
\end{equation}
In two recent papers Naz et al. $(2014, 2016)$ developed a new methodology for solving the dynamical system of first-order ordinary differential equations arising from first-order conditions of optimal control problems. They derived closed-form solutions for the Ramsey model $(1928)$, in their first paper and for the Lucas-Uzawa model $(1965)$, $(1988)$ in the second paper. More recently Naz and Chaudry $(2017)$, give some clarifications on these solutions and made some comparisons with other solutions, previously provided by Boucekkine and Ruiz-Tamarit $(2008)$, Chilarescu $(2011)$ and, Marsiglio and La Torre $(2012)$. As it is well-known for the two models there are some other papers which have found closed-form solutions like those of Barro and Sala-i-Martin $(2004)$, Smith $(2006)$,  Ragni {\it et al.} $(2012)$, Viasu $(2014)$, Hiraguchi $(2009)$ and, Chilarescu and Viasu $(2016)$.

Naz and Chaudry obtained three first integrals, denoted by $I_1,\;I_2$ and $I_3$, the first two with no restrictions on parameters and the last one with a restriction on parameters. Among the two first integrals, only $I_1$ enables us to obtain directly the solutions for the Lucas model. It is impossible to obtain solutions for the Lucas model by using only the second integral $I_2$. That is why it is necessary to combine the two first integrals $I_1$ and $I_2$ in order to obtain the solutions. The solutions thus obtained for the variables $k$ and $c$, coincide exactly with those of the previous case, but the solutions for the variables $h$ and $u$ do not coincide with those of the previous case.

If the solution for $u$ is really a new solution, then the authors have to prove that it is an admissible solution, i.e. $0 < u < 1$ and then they have to determine the starting values for the control variables $u$ and $c$, because these initial values are unknown. None of these requirements can be found in the papers of the cited authors. In the next section we will prove that the Lucas-Uzawa model admits a unique solution and thus the claim of Naz and Chaudry on the existence of multiple solutions is inexact.

\section{The unique solution for the model of Lucas-Uzawa}\label{s:2}
In order to solve the system \eqref{DS}, Chilarescu introduced the new variable $z=\frac{hu}{k}$ and thus he obtained the following differential equation
$$\dot{z} = \left[\frac{\delta+\pi}{\beta}
-\gamma z^{1-\beta}\right]z,$$ whose solution is given by
\begin{equation}\label{eqsolz}
z(t)= \left[\frac{z^{1-\beta}_*z^{1-\beta}_0}{\left(z^{1-\beta}_*-z^{1-\beta}_0\right)e^{-\varphi t}+z^{1-\beta}_0}\right]^{\frac{1}{1-\beta}}.
\end{equation}
As was proved by Chilarescu and by Naz and Chaudry, the solutions for $k$ and $c$ of the system \eqref{DS} are given by
\begin{equation}\label{eqsolc}
c(t)=\frac{h_0u_0}{A_*}\left[z(t)\right]^{-\frac{\beta}{\sigma}}e^{\frac{\delta-\rho}{\sigma}t},
\end{equation}
\begin{equation}\label{eqsolk}
k(t) =  \frac{h_0u_0}{A_*} \left[A_* - A(t)\right]\left[z(t)\right]^{-1}e^{\phi t},\;\phi=\frac{\delta+\pi(1-\beta)}{\beta},
\end{equation}
\begin{equation}\label{eqsolck}
\frac{c(t)}{k(t)} = \frac{\left[z(t)\right]^{\frac{\sigma-\beta}{\sigma}}e^{-\xi t}}{A_* - A(t)},\;\xi=\phi-\frac{\delta-\rho}{\sigma},
\end{equation}
where
$$A(t)=\int\limits_0^t z(s)^{\frac{\sigma-\beta}{\sigma}}e^{-\xi s}ds,\;A_* = \lim\limits_{t\rightarrow\infty}A(t).$$
Substituting \eqref{eqsolck} into the last equation of the system \eqref{DS} we arrive at the following nonlinear differential equation
$$\dot{u} = \left[\varphi - \frac{z^{\frac{\sigma-\beta}{\sigma}}e^{-\xi t}}{A_*-A(t)}
+\delta u\right]u.$$
As was proved by Chilarescu, the starting value $u_0$ can be determined and is the unique solution of the equation $$\left(\varphi +\delta u_0\right) A_*(u_0;k_0,h_0) - \delta u_0 B_*(u_0;k_0,h_0) = 0.$$
Consequently, since the function
$$F(t,u) = \left[\varphi - \frac{z^{\frac{\sigma-\beta}{\sigma}}e^{-\xi t}}{A_*-A(t)}
+\delta u\right]u,$$
is continuously differentiable, than via the existence and uniqueness theorem for nonlinear differential equations, there exists one and only one solution to the initial value problem
$$\dot{u} = F(t,u), \;u_0=u(0).$$
This solution is given by
\begin{equation}
u(t) =\frac{\varphi u_0[A_*-A(t)]}{\left[\left(\varphi+\delta u_0\right)A_*-\delta u_0B(t)\right]e^{-\varphi t} -\delta u_0[A_*-A(t)]},
\end{equation}
where
$$B(t) = \int\limits_0^t z(s)^{\frac{\sigma-\beta}{\sigma}}e^{-\left(\xi - \varphi\right)s}ds,\;B_* = \lim\limits_{t\rightarrow\infty}B(t).$$
Therefore, the claim of Naz et al. and Naz and Chaudry concerning the existence of multiple solutions for the Lucas-Uzawa model is inexact.

In fact, the second set of solutions determined by Naz and Chaudry is identical to that in the first case, but is only written in a different mathematical formulation. The solutions for the state variable $k$ and for the control variable $c$ coincide exactly to those determined in the first set of solutions (the same solutions were determined by Chilarescu). Only the solutions for the state variable $h$ and for the control variable $u$ were determined in a different mathematical formulation. These solutions, written in accordance with the notations used in this paper, are:
\begin{equation}\label{eqhn}
h(t) = \frac{\left\{\left[A_*-A(t)\right]\left[\gamma\beta(1-\sigma)-
(\rho+\pi-\pi\sigma)z(t)^{\beta-1}\right]+\sigma z(t)^{\beta-\frac{\beta}{\sigma}}e^{-\xi t}\right\}e^{\phi t}}{\delta^{-1}\gamma(1-\beta)(\rho-\delta+\delta\sigma)c_0^{-1}z_0^{-\frac{\beta}{\sigma}}},
\end{equation}
\begin{equation}\label{equn}
u(t) = \frac{\delta^{-1}\gamma(1-\beta)(\rho-\delta+\delta\sigma)\left[A_*-A(t)\right]}{\left[A_*-A(t)\right]\left[\gamma\beta(1-\sigma)-
(\rho+\pi-\pi\sigma)z(t)^{\beta-1}\right]+\sigma z(t)^{\beta-\frac{\beta}{\sigma}}e^{-\xi t}}.
\end{equation}
If in the paper of Chilarescu, we express the function $B$ in terms of the function $A$, i.e.,
$$B(t) = \frac{\varphi+\delta u_0}{\delta u_0}A_*-\left[A_*-A(t)\right]\left[1+\mu -\chi z(t)^{\beta-1}\right]e^{\varphi t} - \omega z(t)^{\beta-\frac{\beta}{\sigma}}e^{-(\xi-\varphi)t},$$
with
$$\mu=\frac{\gamma\beta\varphi(1-\sigma)}{\eta},
\chi=\frac{\varphi\left[\rho+\pi(1-\sigma\right]}{\eta},
\omega=\frac{\sigma\varphi}{\eta},\eta=\gamma(1-\beta)\left[\rho-\delta(1- \sigma)\right].$$
and then substitute this result into the corresponding equations of $u$ and $h$ given in the paper of Chilarescu, we obtain exactly the same results as those of the equations obtained by Naz and Chaudry.

The cited authors also claim that under the specific restriction
\begin{equation}\label{eqsolsigma}
\sigma=\frac{(\rho+\pi)\beta}{\pi\beta-(\delta+\pi)(1-\beta)}=
\frac{1}{\frac{\pi}{\rho+\pi}-\frac{\delta+\pi}{\rho+\pi}\frac{1-\beta}{\beta}},
\end{equation}
there exists another solution of the Lucas-Uzawa model. The problem here is that we cannot choose arbitrarily the values of all parameters in the model of Lucas. As it is well-known, $\beta$ represents the capital share of income. For example, if we choose $\pi = 0.05$, $\delta = 0.06$, $\rho = 0.04$ which are acceptable values, then we have to choose (for example) $\beta = 0.8$, value that generates for $\sigma = 4$. These two values, for $\beta$ and $\sigma$ are certainly beyond the values confirmed by the econometric estimations and consequently this second solution could be considered only as a purely mathematical alternative.

In the next section we present some numerical simulations in order to show that the trajectories determined by Naz et al. or by Naz and Chaudry (equations \eqref{eqhn} and \eqref{equn}), coincide exactly with those provided by Chilarescu.
\section*{Conclusions and some numerical simulations}
The uniqueness of the solutions of Lucas's model was proved for the first time by Boucekkine and Ruiz-Tamarit and later by Chilarescu, by using completely different mathematical techniques. Naz et al. and Naz and Chaudry recently published several papers in which they claim that Lucas's model, without any restrictions on the parameters, presents multiple solutions. Obviously this claim is not true and this paper clarifies definitively this subject. We proved our result, via the theorem of existence and uniqueness of the nonlinear differential equations. Examining the results presented in the papers by these authors, results obtained via the partial Hamiltonian approach, we conclude that this method could not provide new general results, but only to confirm the old results obtained by other papers. What is really new to this method is the fact that it can produce some particular solutions, obtained by using different restrictions on the parameters. In order to give more credibility to the results obtained in this paper, we present below, some numerical simulations. To do this, we consider here a well-known benchmark economy:
$\beta=0.25, \gamma=1.05, \delta=0.05, \pi=0.01, \rho=0.04, \sigma=1.5, h_0=10,$
and $k_0=80$ and the results are presented into the four graphs, denoted by $Fig. No. 1 - Fig. No. 4$.
\begin{center}
\begin{figure}[htbp]
\includegraphics[width=5.5cm]{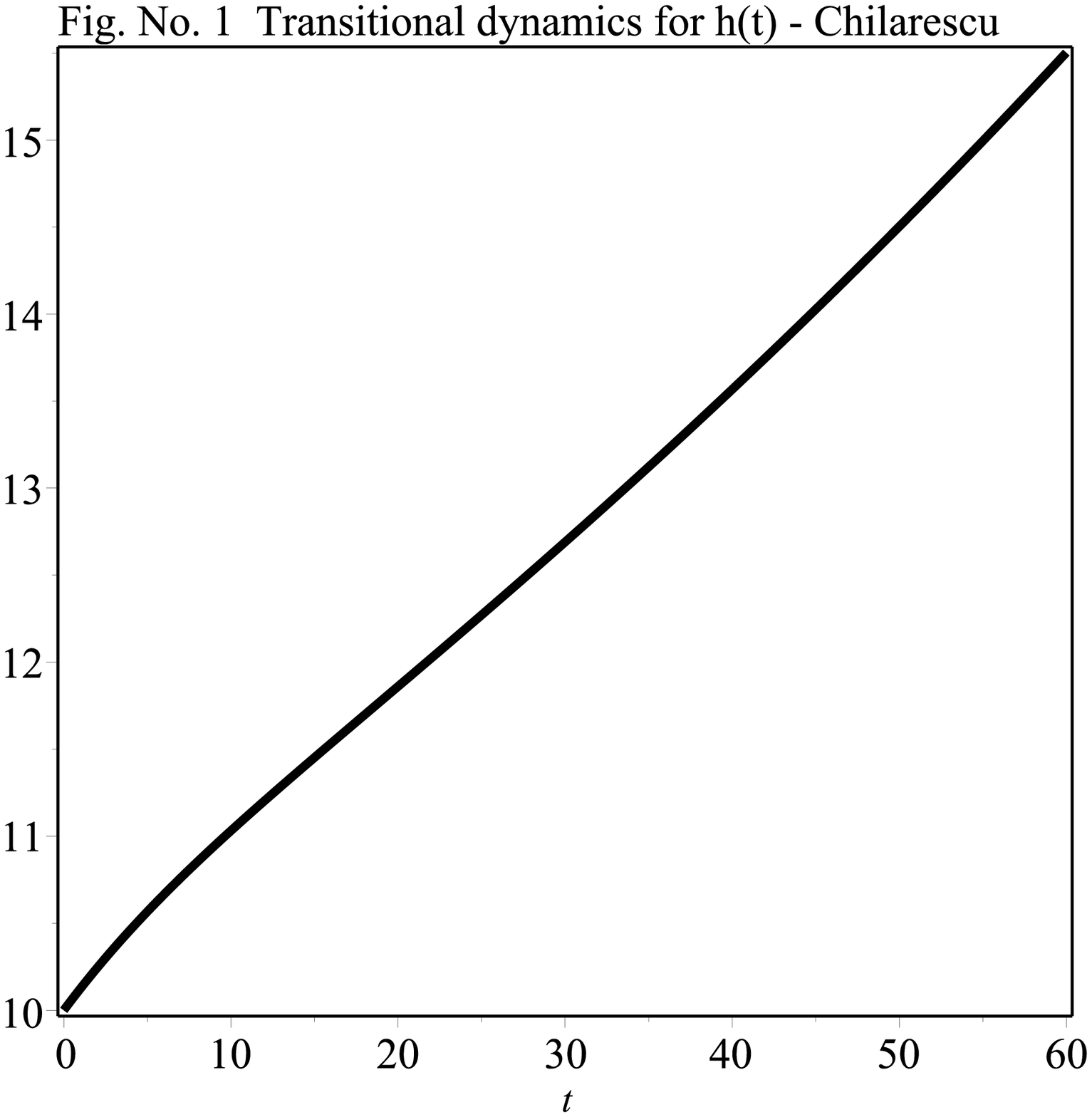}\;\;\includegraphics[width=5.5cm]{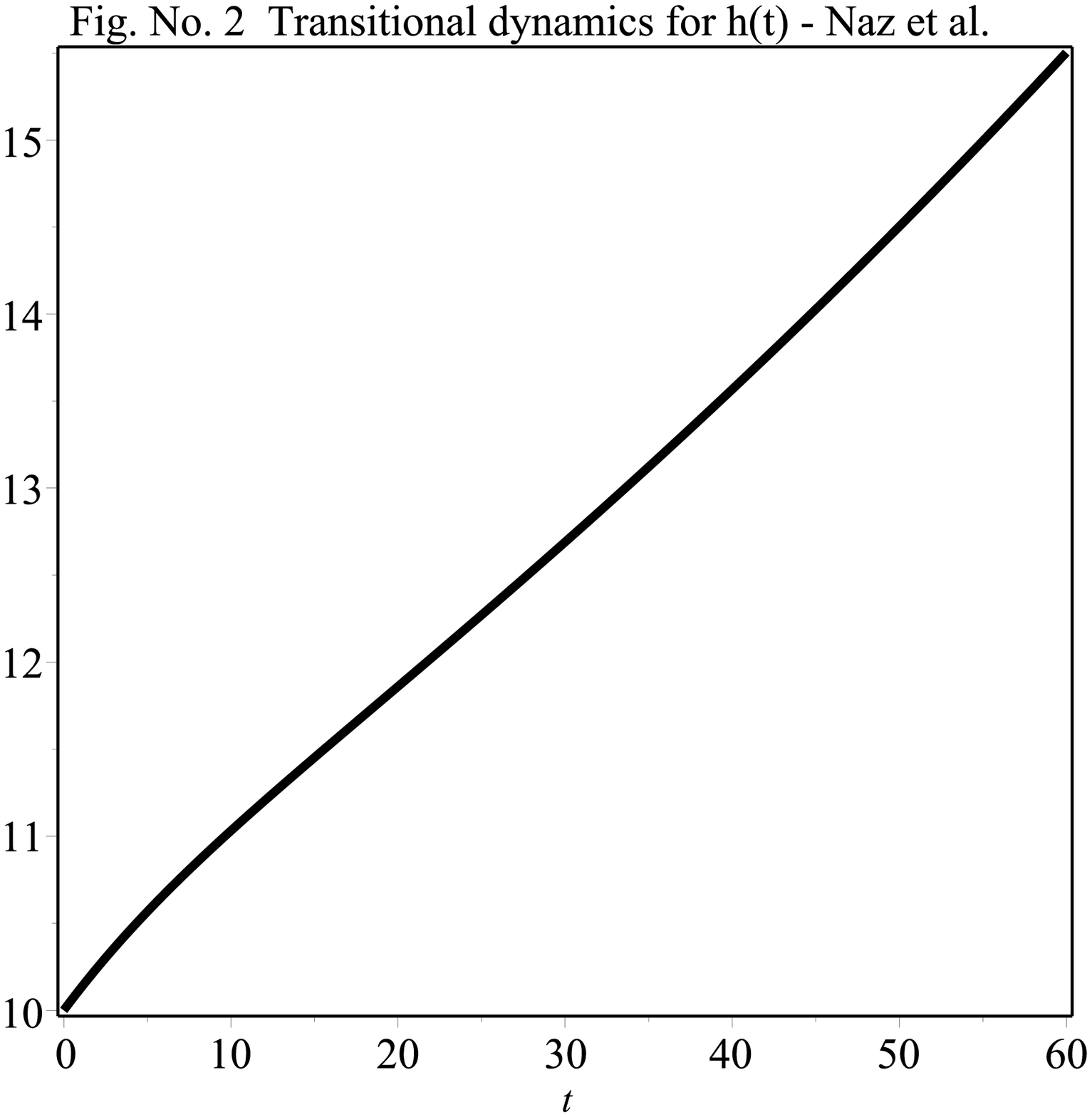}
\end{figure}
\end{center}
\begin{center}
\begin{figure}[htbp]
\includegraphics[width=5.5cm]{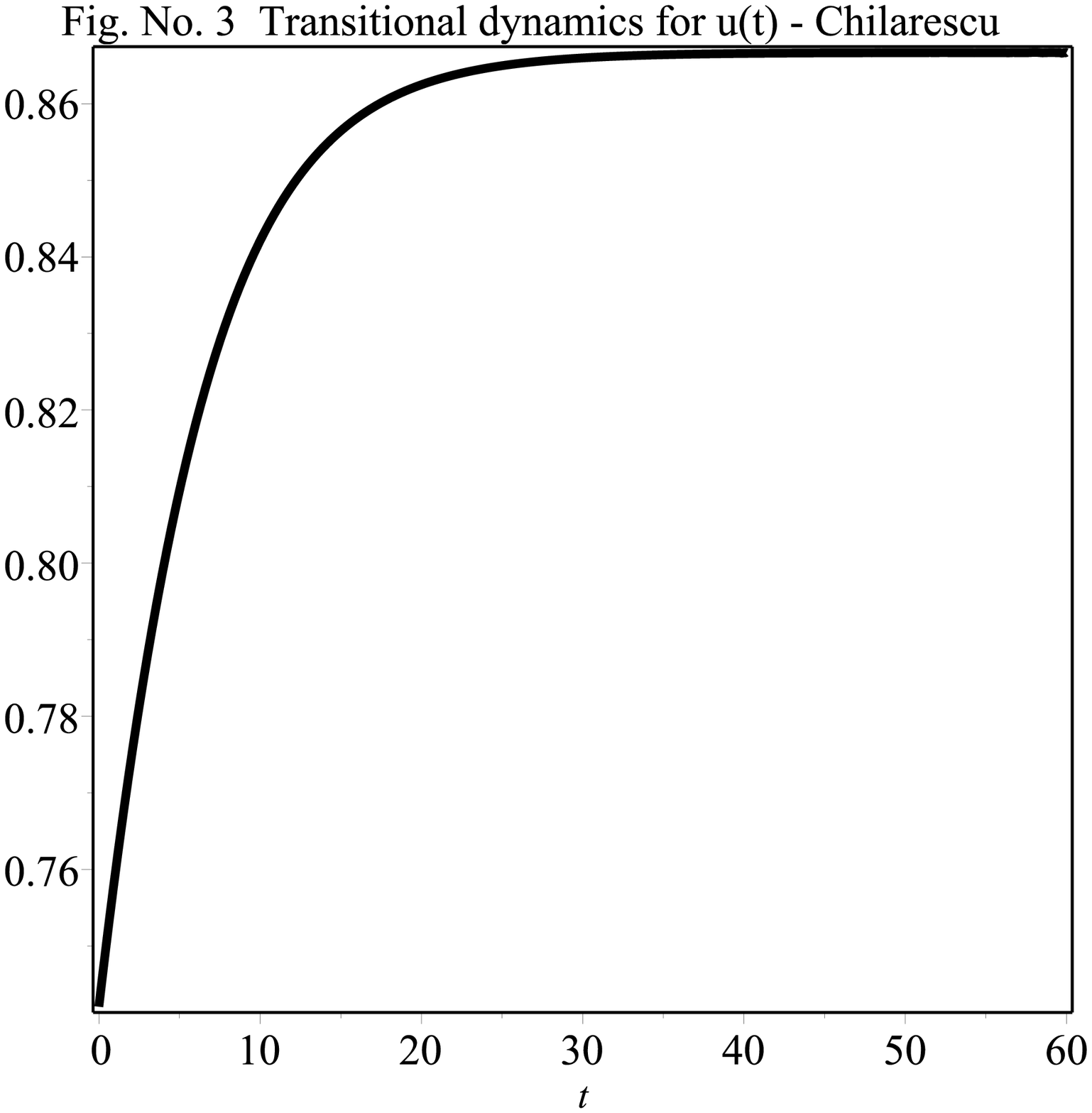}\;\;\includegraphics[width=5.5cm]{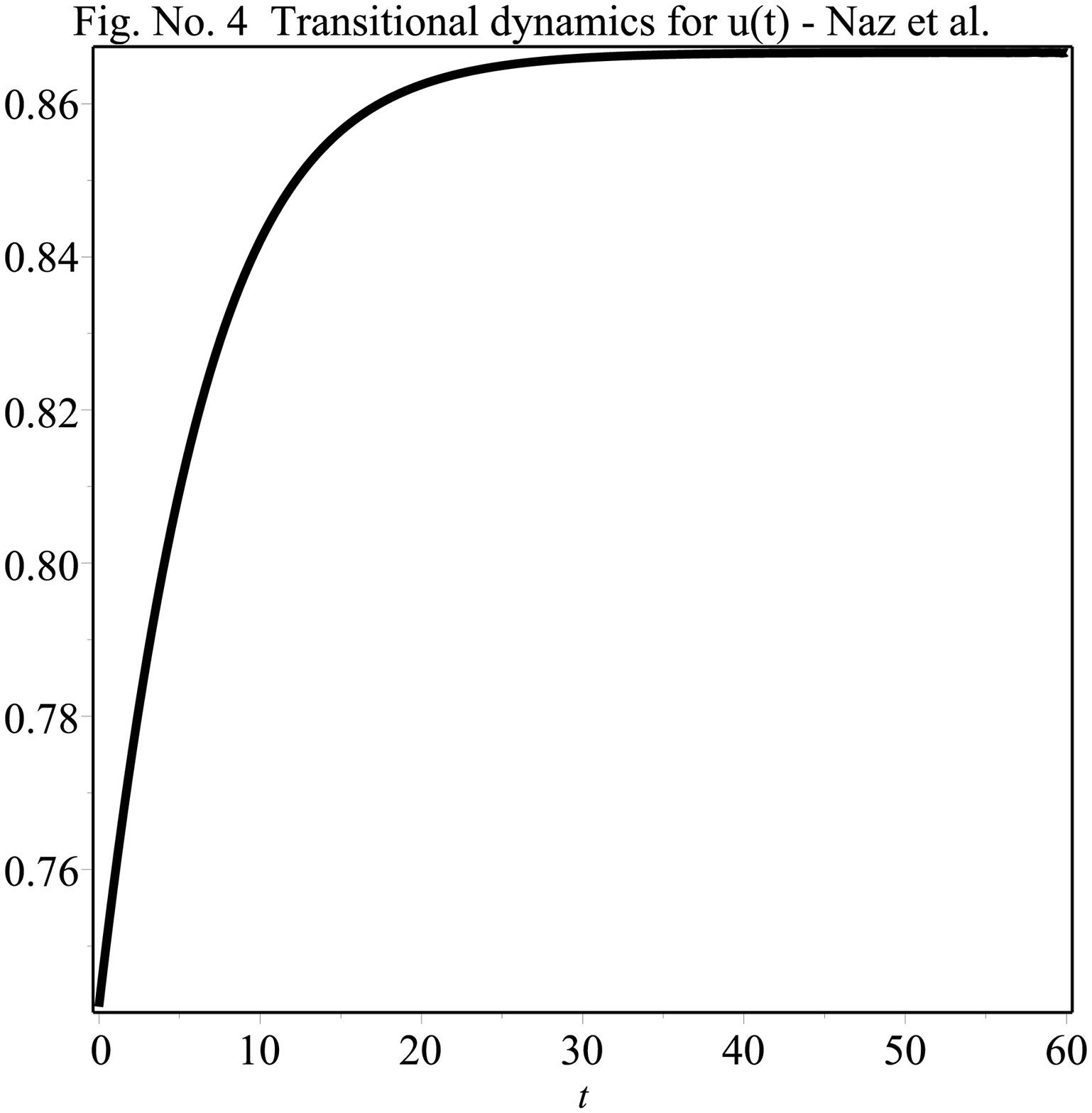}
\end{figure}
\end{center}
As we can observe from the these graphs, the trajectories for the variables $h$ and $u$ obtained by Naz et al. coincide exactly with those obtained by Chilarescu.

\end{document}